\documentclass[a4paper,11pt]{article}
\usepackage{jcappub}
\arxivnumber{2007.09063}

\providecommand\inspire[1]{\href{https://inspirehep.net/search?p=find+#1}{{\tiny IN}{\footnotesize SPIRE}}}

\providecommand{\jhep}[3] {\ifnum#2>2009%
\href{https://doi.org/10.1007/JHEP#1(#2)#3}{\emph{JHEP} {\bf #1} (#2) #3}%
\else%
\href{https://doi.org/10.1088/1126-6708/#2/#1/#3}{\emph{JHEP} {\bf #1} (#2) #3}%
\fi}
\providecommand{\jcap}[3] {\href{https://doi.org/10.1088/1475-7516/#2/#1/#3}{\emph{JCAP} {\bf #1} (#2) #3}} 
\def\issueFromCounter.#1#2#3#4#5#6.{#2#3}
\providecommand{\jstat}[2]{\PackageWarningNoLine{\jname}{The macro \protect\jstat\space is obsolete!\MessageBreak Please typeset JSTAT as any other journal}%
  \href{https://doi.org/10.1088/1742-5468/#1/\issueFromCounter.#2./#2}{\emph{J.\ Stat.\ Mech.\ }(#1) #2}} 

\providecommand{\astroph}[1]{\href{https://arxiv.org/abs/astro-ph/#1}{\tt astro-ph/#1}}

\providecommand{\arXivid}[1]{\href{https://arxiv.org/abs/#1}{\tt arXiv:#1}}
\providecommand{\Math}[2]{%
\if!#1!%
\href{https://arxiv.org/abs/math/#2}{\tt math/#2}%
\else%
\href{https://arxiv.org/abs/math.#1/#2}{\tt math.#1/#2}%
\fi}

\usepackage{bm}

\title{Cosmic ray propagation in the Universe in presence of a random magnetic field}

\author{A.D.~Supanitsky}

\affiliation{Instituto de Tecnolog\'ias en Detecci\'on y Astropart\'iculas (CNEA, CONICET, UNSAM),\\ Centro At\'omico Constituyentes,\\ San Mart\'in, Buenos Aires, Argentina}

\emailAdd{daniel.supanitsky@iteda.cnea.gov.ar}

\abstract{The origin of the ultrahigh energy cosmic ray remains being a mystery. However, a
considerable progress has been made in the past few years due to the good quality data recorded
by current cosmic ray observatories. One of the recent achievements is obtaining firm observational
evidence about the extragalactic origin of the most energetic cosmic rays by the Pierre Auger
observatory. On the other hand, it is believed that there is a non-null turbulent magnetic field
that fills the intergalactic medium. Therefore, the presence of the intergalactic magnetic field
can play an important role on the propagation of the ultrahigh energy cosmic rays through the
Universe, which in principle can be relevant to interpret the experimental data. In this work we
present a system of partial differential equations that describes the propagation of the ultrahigh
energy cosmic rays through the Universe, in the presence of a turbulent intergalactic magnetic field,
that includes the diffusive and the ballistic regime of propagation and also the transition between
them. Also, as an example of application, the system of equations is solved numerically in a
simplified physical situation.}

\begin{document}
\maketitle
\flushbottom

\section{Introduction}
\label{Int}

Despite the significant progress made in recent years, especially in the field of experimental
research, the origin and nature of the ultrahigh energy cosmic rays (UHECRs, $E\geq 10^{18}$\,eV)
are still unknown. The Pierre Auger Observatory, located in the southern hemisphere, and Telescope
Array, located in the northern hemisphere, are currently taking data of very good quality. Also
the current statistics is quiet large, specially the one corresponding to Auger which at present
has accumulated a very large exposure.

The UHECR flux has been measured with unprecedented statistics by Auger and Telescope Array. It
presents two main features, a hardening at $\sim 10^{18.7}$\,eV, known as the ankle and a suppression.
This suppression is observed by Auger at $10^{(19.62 \pm 0.02)}$\,eV and by Telescope Array at a larger
energy, $10^{(19.78 \pm 0.06)}$\,eV~\cite{AugerTA:17}. Also, the Auger spectrum takes smaller values
than the ones corresponding to Telescope Array. The discrepancies between the two observations can
be diminished by shifting the energy scales of both experiments within their systematic uncertainties.
However, some differences are still present in the suppression region~\cite{AugerTA:17}.

It has long been believed that the most energetic part of the UHECR flux is of extragalactic
origin. This is due to the known inefficiencies of the galactic sources to accelerate particles
at the highest energies. Moreover, very recently Auger has found strong observational evidence
about the extragalactic origin of the cosmic rays with primary energies above $\sim 10^{18.9}$\,eV
from the study of the distribution of their arrival directions. Considering this data set an
anisotropy that can be described as a dipole of $\sim 6.5$\% amplitude was found~\cite{Science:17}.
The significance of this detection is larger than $5.2\, \sigma$ and the dipole direction is such
that a scenario in which the flux is dominated by a galactic component is disfavored~\cite{Science:17}.

The transition between the galactic and extragalctic components is still an open problem of the
high energy astrophysics. The Auger data show that the large scale distribution of the cosmic ray
arrival directions is compatible with an isotropic flux, in the energy range from $\sim 10^{18}$\,eV
up to the ankle~\cite{Auger:12}. This result is incompatible with a galactic origin of the light
component that seems to dominate the flux in this energy range~\cite{Auger:12}. Therefore, the
transition between these two components is expected to take place below $10^{18}$\,eV.

Thus, the large scale arrival direction analyses done by Auger, combined with the composition
information at lower energies, shows that the UHECR flux seems to be dominated by the extragalactic
component.

It is believed that the intergalactic medium is filled with a turbulent magnetic field, which
affects the propagation of the charged extragalactic UHECRs through the Universe. In particular,
the effects produced on the propagation of charged nuclei are relevant to explain the UHECR data
(see refs.~\cite{Mollerach:19,Mollerach:20} for recent works).

The propagation of charged particles in a random magnetic field depends on the distance traveled
by the particles under the influence of the field compared with the scattering length
$\lambda_{SL}=3D/c$, where $D$ is the diffusion coefficient and $c$ is the speed of light. For
traveled distances much smaller than $\lambda_{SL}$, which in general corresponds to the region
close to the source, the propagation is ballistic. For traveled distances much larger than
$\lambda_{SL}$, the propagation is diffusive (see, for instance, ref.~\cite{Harari:14}).

The propagation of the UHECRs in the presence of the intergalactic magnetic field can be
study from simulations~\cite{CRPropa3:16} or through the appropriated transport equation that
describes this physical problem. In ref.~\cite{Berezinsky:06} a partial differential equation
is obtained for the number density of particles which is valid for the diffusive regime of
propagation. This equation can be solved analytically provided the diffusion coefficient for
a given random magnetic field model. A phenomenological extension of the solution found in ref.~\cite{Berezinsky:06} to the ballistic case is proposed in ref.~\cite{Aloisio:09}.

On the other hand, in ref.~\cite{Prosekin:15} a system of partial differential equations is obtained
applying the method of moments to the Boltzmann equation with an appropriated collision term
introduced to describe the propagation of cosmic rays generated in galactic sources, for which the
effects of the expansion of the Universe are negligible. In this case the equations system is formed
by two coupled partial differential equations for the number density of particles and the flux.
These equations take into account the ballistic and the diffusive regimes of propagation and
also the transition between both of them. Also, an analytic solution of this equations system is
found for the stationary case. In this work we find a system of partial differential equations
also for the number density of particles and the flux applying the methods of moments to the
Boltzmann equation, which includes both, the effects of the intergalactic magnetic field and the
expansion of the Universe. In this case, the starting point is the Boltzmann equation in a curved
space-time. We also solve numerically the equations system for a simplified physical situation
which shows that the equations found properly describe the ballistic and the diffusive regime of
propagation and also the transition between them.

\section{The transport equation and the method of moments}
\label{TEqs}

The propagation of the cosmic rays in the expanding Universe is described by the Boltzmann
equation in a curved space-time. The Friedmann-Lema\^{i}tre-Robertson-Walker (FLRW) metric
for a spatially flat Universe is considered,
\begin{equation}
\label{FLRWm}
ds^2=-dt^2 + a^2(t) \delta_{ij} dx^i dx^j,
\end{equation}
where $\{x^0,x^1,x^2,x^3\}$ are comoving coordinates with $x^0=t$, $a(t)$ is the scale factor,
and $\delta_{ij}$ is the Kronecker delta function with $i,j = \{1,2,3\}$. In eq.~(\ref{FLRWm})
and in the rest of the article the speed of light $c=1$ is considered.

The geodesic equation is given by,
\begin{equation}
\frac{dP^\mu}{d\lambda}+\Gamma_{\nu \rho}^\mu P^\nu P^\rho = 0,
\end{equation}
where $P^\mu=dx^\mu/d\lambda$ is the four-momentum of the particle, $\lambda$ is the affine
parameter, and $\Gamma_{\nu \rho}^\mu$ are the Christoffel symbols. For the FLRW metric the
non-null Christoffel symbols are: $\Gamma_{ij}^0 = a(t) \dot{a}(t) \delta_{ij}$ and
$\Gamma_{0j}^i = H(t) \delta_{j}^i$, where $\dot{a}(t)=da(t)/dt$ and $H(t)=\dot{a}(t)/a(t)$
is the Hubble parameter. The geodesic equation for the spatial components of the four-momentum
and for the FLRW metric takes the following form,
\begin{equation}
\frac{dP^i}{d\lambda}+2 H(t) P^0 P^i = 0.
\label{dPdl}
\end{equation}

It is appropriated to work in the \emph{local inertial frame}, which corresponds to an observer
placed at a point with $x^\mu$ space-time coordinates~\cite{Pettinari:16}. It is worth mentioning
that in this frame the collision term of the Boltzmann equation takes the same form as in Special
Relativity. The tangent space associated to a given space-time point is spanned by a basis of four
contravariant vectors which are called the \emph{tetrad}. The choice of the \emph{tetrad}
defines the reference frame in that point. The natural \emph{tetrad} is the one known as the
coordinate \emph{tetrad}, which corresponds to the directional derivatives with respect to the
coordinates. The \emph{tetrad} corresponding to the \emph{local inertial frame} is the one in
which the metric takes the form of the one corresponding to the Minkovski space-time. Therefore,
the four-momentum in the \emph{local inertial frame} takes the following form (see for instance ref.~\cite{Berenstein:88}),
\begin{equation}
\begin{aligned}
p^0 &= P^0 \\
p^i &= a(t) P^i
\end{aligned}
.
\label{pLIF}
\end{equation}
Note that in the \emph{local inertial frame} the relationship between the energy of a particle
and the momentum is $E=\sqrt{p^2 + m^2}$, where $p = \sqrt{(p^1)^2+(p^2)^2+(p^3)^2}$. For massless
particles or in the case where $E \gg m$, $E=p$.

The equation fulfilled by the spatial components of the four-momentum in the \emph{local inertial frame}
is obtained from eqs.~(\ref{dPdl}) and~(\ref{pLIF}),
\begin{equation}
\frac{dp^i}{d\lambda} + H(t) p^0 p^i = 0.
\label{dpdl}
\end{equation}
where $p^0 = E$.

The Boltzmann equation in a curved space-time is given by~\cite{Pettinari:16},
\begin{equation}
\frac{d f}{d\lambda} = C[f] + S,
\label{BE0}
\end{equation}
where $C[f]$ is the collision term and $S$ is a source term. Here the function $f$ depends on the
space-time coordinates of a point and on the spatial coordinates of the four-momentum in the
\emph{local inertial frame}, i.e.~$f \equiv f(t,\mathbf{x},\mathbf{p})$ where
$\mathbf{x}=(x^1,x^2,x^3)$ and $\mathbf{p}=(p^1,p^2,p^3)$.

By using eq.~(\ref{dpdl}) the Boltzmann equation becomes,
\begin{equation}
\label{BE}
\frac{\partial f}{\partial t} + \frac{\mathbf{\hat{p}}}{a} \cdot \bm{\nabla}_{\mathbf{x}} f %
-H E\ \mathbf{\hat{p}} \cdot \bm{\nabla}_{\mathbf{p}} f = C[f] + S,
\end{equation}
where $\mathbf{\hat{p}}=\mathbf{p}/p$ is a unit vector pointing in the direction of motion of the
particles,
$\bm{\nabla}_{\mathbf{x}} = (\partial/\partial x^1, \partial/\partial x^2, \partial/\partial x^3)$,
and $\bm{\nabla}_{\mathbf{p}} = (\partial/\partial p^1, \partial/\partial p^2, \partial/\partial p^3)$.

The collision term considered is the one introduced in ref.~\cite{Prosekin:15},
\begin{equation}
C[f]=\int d\Omega'\, \omega(t,\mathbf{x},E,\mathbf{\hat{p}'},\mathbf{\hat{p}})
f(t,\mathbf{x},E\, \mathbf{\hat{p}'})-
\int d\Omega'\, \omega(t,\mathbf{x},E,\mathbf{\hat{p}},\mathbf{\hat{p}'}) f(t,\mathbf{x},E\, \mathbf{\hat{p}}),
\end{equation}
where $\omega(t,\mathbf{x},E,\mathbf{\hat{p}_1},\mathbf{\hat{p}_2})$ is the scattering probability for a
particle with initial direction of motion $\mathbf{\hat{p}_1}$ and final direction of motion $\mathbf{\hat{p}_2}$
per unit of time at the point $\mathbf{x}$. Note that it assumed that $\mathbf{p}=E\, \mathbf{\hat{p}}$, which
corresponds to the case where $E \gg m$, a very good approximation for UHECRs.

The source term considered is the following,
\begin{equation}
S = s(t,\mathbf{p}) \delta(\mathbf{x}-\mathbf{x}_s) \Theta(t-t_g),
\end{equation}
which corresponds to a point source placed at $\mathbf{x}_s$ that started to emit cosmic rays at $t_g$.
Here $\delta(\mathbf{x})$ is the delta Dirac function and $\Theta(t)$ is the Heaviside function.

The four-current in the \emph{local inertial frame} is given by~\cite{Berenstein:88},
\begin{equation}
N^a(t,\mathbf{x}) = \int d^3p\, \frac{ p^a }{E} f(t,\mathbf{x},\mathbf{p}).
\end{equation}
Therefore, the number density of particles and the flux differential in energy are given by,
\begin{align}
\label{Den}
n(t,\mathbf{x},E) &= E^2 \int d\Omega\, f(t,\mathbf{x},E\, \mathbf{\hat{p}}),\\
\label{Flux}
\mathbf{J}(t,\mathbf{x},E) &= E^2 \int d\Omega\, \mathbf{\hat{p}} f(t,\mathbf{x},E\, \mathbf{\hat{p}}).
\end{align}

Following ref.~\cite{Prosekin:15}, after applying the integral operators $\int d\Omega$ and
$\int d\Omega\, \mathbf{\hat{p}}$ to eq.~(\ref{BE}), the equations for the number density of
particles and the flux are obtained,
\begin{align}
\label{n}
&\frac{\partial n}{\partial t} - \frac{\partial}{\partial E}(b\, n) + 3 H n +%
\frac{1}{a} \bm{\nabla} \cdot \mathbf{J} = q(t,E) \delta(\mathbf{x}-\mathbf{x}_s) \Theta(t-t_g), %
\\
\label{J}
&\frac{\partial J^i}{\partial t} - \frac{\partial}{\partial E}(b\, J^i) + 3 H J^i +
\frac{1}{a} \frac{\partial}{\partial x^j} \left( \langle \hat{p}^i \hat{p}^j \rangle n \right)
+ \frac{J^i}{\tau} = 0,
\end{align}
where $b(t,E) = H(t) E + b_{\mathrm{int}}(t,E)$, in which the term $b_{\mathrm{int}}(t,E)$ is included to take into
account processes modeled as continuous energy losses of the particles. Here,
\begin{align}
\label{pipj}
\langle \hat{p}^i \hat{p}^j \rangle(t,\mathbf{x},E) &= %
\frac{{\mathop{\displaystyle \int d\Omega\, \hat{p}^i \hat{p}^j %
f(t,\mathbf{x},E\, \mathbf{\hat{p}})}}}{{\mathop{\displaystyle \int d\Omega f(t,\mathbf{x},E\,
\mathbf{\hat{p}})}}}, \\
\label{Tau}
\tau(t,\mathbf{x},E) &= \left[\int d\Omega (1-\mathbf{\hat{p}}\cdot \mathbf{\hat{p}'})\,%
\omega(t,\mathbf{x},E,\mathbf{\hat{p}},\mathbf{\hat{p}'})  \right]^{-1},
\end{align}
where $\langle \hat{p}^i \hat{p}^j \rangle$ is the isotropization tensor. Note that these expressions
are the same as the ones obtained in ref.~\cite{Prosekin:15}. The assumptions to obtain eqs.~(\ref{n}),~(\ref{J}),~(\ref{pipj}), and~(\ref{Tau}) are: $\int d\Omega\, C[f] = 0$, the source term does not
depends on $\mathbf{\hat{p}}$, i.e.~$s(t,\mathbf{p}) = s_I(t,E)$ which corresponds to a source
emitting cosmic rays isotropically, and $q(t,E)= 4\pi\, s_I(t,E)$.

It is worth mentioning that, the number density of particles $n$ and the flux $\mathbf{J}$ are
related to observable physical quantities~\cite{Ahlers:17} (see also refs.~\cite{Ahlers:14} and~\cite{Deligny:19}). Following ref.~\cite{Ahlers:17}, the cosmic ray intensity can be written as,
\begin{equation}
I(t,\mathbf{x},E,\mathbf{\hat{p}}) = E^2 \, f(t,\mathbf{x},-E\mathbf{\hat{p}}) = \frac{1}{4 \pi} n(t,\mathbf{x},E) - %
\frac{3}{4 \pi} \mathbf{J}(t,\mathbf{x},E)\cdot \mathbf{\hat{p}} + \mathcal{O}(\{a_{lm}\}_{l \geq 2}),
\end{equation}
where $\mathbf{\hat{p}}$ points to a given direction in the sky and $\mathcal{O}(\{a_{lm}\}_{l \geq 2})$
corresponds to terms with $l \geq 2$ of the multipole expansion. Therefore, $n$ is proportional to the
average cosmic ray intensity over the entire sky and $\mathbf{J}$ is related to the dipole vector, which
is defined as~\cite{Ahlers:17},
\begin{equation}
\label{DipoleV}
\bm{\Delta} (t,\mathbf{x},E) = -3\, \frac{\mathbf{J}(t,\mathbf{x},E)}{n(t,\mathbf{x},E)}.
\end{equation}
The dipole amplitude is given by the norm of the dipole vector, i.e.\ $\Delta = ||\bm{\Delta}||$. The
dipole vector can be reconstructed from the distribution of the arrival directions of the UHECRs~\cite{Science:17}.

In the diffusive regime the first three terms of eq.~(\ref{J}) can be discarded and also
since $\langle \hat{p}^i \hat{p}^j \rangle = \delta^{ij}/3$, the following expression for the
flux as a function of the number density of particles is obtained,
\begin{equation}
\label{Jdiff}
\mathbf{J}(t,\mathbf{x},E) = -\frac{D(t,\mathbf{x},E)}{a(t)}\, \bm{\nabla} n(t,\mathbf{x},E),
\end{equation}
where $D(t,\mathbf{x},E)=\tau(t,\mathbf{x},E)/3$ is the diffusion coefficient. Introducing
eq.~(\ref{Jdiff}) in eq.~(\ref{n}) the equation for the number density of particles of ref.~\cite{Berezinsky:06} is obtained. Therefore, the number density of particles obtained
solving eqs.~(\ref{n}) and~(\ref{J}) has to be approximately equal to the analytic solution
found in ref.~\cite{Berezinsky:06} when the diffusive limit is considered.

Note that, the equations system of ref.~\cite{Prosekin:15} is obtained introducing $a(t)=1$
and $b_{\mathrm{int}}(t,E) = 0$ in equations~(\ref{n}) and~(\ref{J}). Also note that the source term
in eq.~(\ref{n}) is different than the one of ref.~\cite{Prosekin:15} due to the different
physical scenario considered in this work.

Let us consider the case in which the cosmic ray source is at the coordinates origin and
that there is spherical symmetry. Under these assumptions the number density of particles and
the flux depend only on the radial coordinate $r=||\mathbf{x}||$ and the flux has a non-null
component in the radial direction, $\mathbf{\hat{r}}=\mathbf{x}/r$, only. Also, following ref.~\cite{Prosekin:15}, it is assumed that the isotropization tensor has the following shape,
\begin{equation}
\langle \hat{p}^i \hat{p}^j \rangle(t,\mathbf{x},E)=(1-\phi(t,r,E)) \frac{\delta^{ij}}{3} + %
\phi(t,r,E) \, \hat{r}^i \hat{r}^j,
\end{equation}
where $\phi(t,r,E)$ is the isotropization function and $\hat{r}^k = x^k/r$. 

The equations for the number density of particles and the radial component of the flux, $J$, are
given by,
\begin{align}
\label{nr}
&\frac{\partial n}{\partial t} - \frac{\partial}{\partial E}(b\, n) + 3 H n +%
\frac{1}{a \, r^2} \frac{\partial }{\partial r}(r^2 \, J) = q \, %
\frac{\delta(r)}{4\pi r^2} \, \Theta(t-t_g), \\
\label{Jr}
&\frac{\partial J}{\partial t} - \frac{\partial}{\partial E}(b\, J) + 3 H J +%
\frac{(1+2\phi)}{3\, a} \frac{\partial n}{\partial r} + %
\frac{1}{a} \left( \frac{2}{3} \frac{\partial \phi}{\partial r} + \frac{2}{r} \phi \right) n +%
\frac{J}{\tau} = 0.
\end{align}

In order to simplify the equations for $n$ and $J$ a change of the variable $E$ to a new
variable $E_0$ is performed. The new variable $E_0$ is obtained by solving the following
equation,
\begin{equation}
\label{bte}
\frac{dE}{dt} = -b(t,E),
\end{equation}
with the condition $E(t_0) = E_0$ where $t_0$ is taken as the age of the Universe. In this way
a function $E(t,E_0,t_0)$ is obtained, which corresponds to the energy that a particle has to have
at time $t$ in order to be observed at time $t_0$ with energy $E_0$. This change of variable is such
that the partial derivative with respect to $E_0$ does not appear in the equations for $n$ and $J$.
Note that this change of variable is the same as the one that appears in the method of characteristics
used to solve systems of partial differential equations~\cite{CharactM}. Therefore, introducing this
change of variable and considering the following functions,
\begin{align}
g(t,r,E_0)\! &= \! 4\pi r^2 \exp\! \left[ \int_{t_g}^{t} dt' \left( 3 H(t') -
\frac{\partial b}{\partial E}(t',E(t',E_0,t_0)) \right) \right]\! n(t,r,E(t,E_0,t_0)), \\
h(t,r,E_0)\! &= \! 4\pi r^2 \exp\! \left[ \int_{t_g}^{t} dt' \left( 3 H(t') -
\frac{\partial b}{\partial E}(t',E(t',E_0,t_0)) \right) \right]\! J(t,r,E(t,E_0,t_0)),
\end{align}
a new equations system for $g$ and $h$ is obtained,
\begin{align}
\label{gr}
&\frac{\partial g}{\partial t} + \frac{1}{a} \frac{\partial h}{\partial r} = %
\exp\left[ \int_{t_g}^{t} dt' \left( 3 H - \frac{\partial b}{\partial E} \right) \right] %
q\, \delta(r)\, \Theta(t-t_g), \\
\label{hr}
&\frac{\partial h}{\partial t} + \frac{(1+2\phi)}{3\, a} \frac{\partial g}{\partial r} + %
\frac{2}{3\, a} \left( \frac{\phi - 1}{r} + \frac{\partial \phi}{\partial r} \right) g + %
\frac{h}{\tau} = 0.
\end{align}
Here the energy variable $E$ in the functions $q$, $\phi$, and $\tau$ is replaced by
the function $E(t,E_0,t_0)$. The functions $g$ and $h$ have to be given as function of the
variable $E$ and not $E_0$. After solving the eqs.~(\ref{gr}) and~(\ref{hr}) the change of
variables from $E_0$ to $E$ has to be done. However, if the solutions are evaluated at time
$t_0$ the change of variable is not necessary, since $E(t_0,E_0,t_0)=E_0$.

The extragalactic magnetic field is poorly known (see ref.~\cite{Han:17} for a review).
Measurements of the magnetic field intensity in galaxy clusters suggest that
the magnetic field intensity in high density regions like sheets and filaments can reach
values up to few $\mu$G. In contrast, observational constraints show that the magnetic
field intensity in voids is smaller than $1-10$ nG. Therefore, the extragalactic magnetic
field can have a strong dependence on $\mathbf{x}$, which is translated into a strong
dependence of the isotropization tensor and the diffusion coefficient on $\mathbf{x}$.
While, the equations system composed by eqs.~(\ref{n}) and~(\ref{J}) describes the general
situation in which both the isotropization tensor and the diffusion coefficient can have
a general dependence on $\mathbf{x}$, the equations system given by eqs.~(\ref{gr}) and~(\ref{hr}) is valid for models of the intergalactic magnetic field with spherical symmetry,
in particular the simple one in which the intergalactic magnetic field is isotropic and
homogeneous.

It is worth mentioning that the equations system obtained can be very useful in the context
of semi-analytical methods used to include the effects of the intergalactic magnetic field in
models of the UHECR flux like the ones developed in refs.~\cite{Mollerach:13,Mollerach:19,Mollerach:20}.

Furthermore, the equations system formed by the $n(t,r,E)$ and $J(t,r,E)$ functions can be
extended in order to include several nuclear species and also the photodisintegration and
photopion production processes undergone by the nuclei during propagation through the radiation
field present in the Universe. This extension, that will be discussed in a forthcoming article,
can be important for the development of models of the UHECR flux that include the effects of
the intergalactic magnetic field. Note that this type of formalism can reduce considerably the
computation time compared to the methods based on Monte~Carlo~simulations.

\section{A simple example}
\label{SE}

Let us consider a simplified case in which a source emits ultrahigh energy protons isotropically,
which lose energy due to the adiabatic expansion of the Universe only (i.e.~$b_{\mathrm{int}}=0$) when they
propagate through the Universe. Let us also consider that the extragalactic magnetic field is
modeled as an isotropic and homogeneous random field such that $\langle \mathbf{B}(\mathbf{x}) \rangle = 0$.
Note that this is a simplified model of the intergalactic magnetic field (see section~\ref{TEqs}).
Therefore, under these assumptions the propagation of the protons is described by eqs.~(\ref{gr}) and~(\ref{hr}). It is also assumed that the source emits cosmic rays at a constant rate in comoving volume
with a spectrum that follows a power law in energy, $q(t,E)= Q(E)/a^3(t)$, where $Q(E)=Q_0 E^{-\gamma}$
with $Q_0$ a constant and $\gamma$ the spectral index.

Since the particles propagate without interacting $b(t,E)= H(t) E$. Introducing this expression in eq.~(\ref{bte}) and solving the differential equation with the corresponding condition, the following
expression for the energy as a function of time and $E_0$ is obtained, $E(t,E_0,t_0) = E_0/a(t)$,
where since $t_0$ is chosen as the age of the Universe it follows that $a(t_0)=1$.

The exponential factor on the right-hand side of eq.~(\ref{gr}) is given by,
\begin{equation}
\exp\left[ \int_{t_g}^{t} dt' \left( 3 H(t') - \frac{\partial b}{\partial E}(t',E(t',E_0,t_0))%
\right) \right] = \frac{a^2(t)}{a^2(t_g)},
\end{equation}
in which it is used that $\partial b/\partial E = H(t)$.

Rescaling the function $g$ and $h$ from eqs.~(\ref{gr}) and~(\ref{hr}) such that
$\xi(t,r,E_0)=g(t,r,E_0)\, a^2(t_g)/Q(E_0)$ and $\eta(t,r,E_0)=h(t,r,E_0)\, a^2(t_g)/Q(E_0)$ the
following equations are obtained,
\begin{align}
\label{xir}
& \frac{\partial \xi}{\partial t} + \frac{1}{a} \frac{\partial \eta}{\partial r} = %
a^{\gamma-1}(t)\, \delta(r)\, \Theta(t-t_g), \\
\label{etar}
& \frac{\partial \eta}{\partial t} + \frac{(1+2\phi)}{3 a} \frac{\partial \xi}{\partial r} + %
\frac{2}{3 a} \left( \frac{\phi - 1}{r} + \frac{\partial \phi}{\partial r} \right) \xi + %
\frac{\eta}{\tau} = 0. \ \ \
\end{align}
Note that for $t=t_0$ these two function are related to $n$ and $J$ through the following expressions,
\begin{align}
\xi(t_0,r,E_0) &= \frac{4\pi r^2}{Q(E_0)}\, n(t_0,r,E_0), \\
\eta(t_0,r,E_0) &= \frac{4\pi r^2}{Q(E_0)}\, J(t_0,r,E_0).
\end{align}
where $\xi$ is the enhancement factor introduced in ref.~\cite{Mollerach:19}. Also, note that
from eq.~(\ref{DipoleV}) it can be seen that the dipole amplitude is given by,
\begin{equation}
\label{DeltaS}
\Delta(t_0,r,E_0) = 3 \, \frac{\eta(t_0,r,E_0)}{\xi(t_0,r,E_0)}.
\end{equation}

In the general case eqs.~(\ref{xir}) and~(\ref{etar}) cannot be solved analytically (an analytic
solution can be found for the ballistic regime where $\phi=1$, see appendix~\ref{AS}). Therefore,
the equations system is solved numerically by using the finite differences method~\cite{Thomas:95}.

The diffusion coefficient used in the numerical calculation is taken from ref.~\cite{Harari:14},
which is given by,
\begin{equation}
D(E)=\frac{l_c}{3} \left[4 \left( \frac{E}{E_c} \right)^2 + 0.9 \left( \frac{E}{E_c} \right) +
0.23 \left( \frac{E}{E_c} \right)^{1/3}   \right],
\end{equation}
where $l_c$ is the coherent length of the random magnetic field. Here $E_c=Z e B\, l_c$, where
$Z$ is the charge number of the nucleus, $e$ is the absolute value of the electron charge, and
$B=\sqrt{\langle \mathbf{B}^2(x) \rangle}$ is the root mean square of the random magnetic field.
This expression corresponds to the type of turbulence given by the Kolmogorov spectrum.

The scale factor and the Hubble parameter considered for the calculation are given by~\cite{KT:94},
\begin{align}
a(t) &= \left( \frac{\Omega_m}{\Omega_\Lambda} \right)^{1/3} %
\sinh^{2/3}\left( \frac{2}{3} \sqrt{\Omega_\Lambda} H_0 t  \right), \\
H(t) &= H_0 \sqrt{\frac{\Omega_m}{a^3(t)}+\Omega_\Lambda},
\end{align}
where $H_0 = 70$ km s$^{-1}$ Mpc$^{-1}$ is the Hubble constant, $\Omega_m = 0.3$ and
$\Omega_\Lambda=0.7$ are the density parameters for matter and dark energy, respectively.

Motivated by ref.~\cite{Prosekin:15}, the isotropization function is assumed to have the following 
expression, $\phi(r,E)= \exp[-\alpha(E)\, r/\tau(E) ]$, where $\alpha(E)$ is a very slowly decreasing
function of energy that ranges from 1.65 at $10^{18}$ eV to 0.935 at $10^{20.5}$ eV. The function 
$\alpha(E)$ is determined in such a way that the numerical solutions $\xi$ and $\eta$ have the form 
corresponding to the ballistic regime of propagation for $r \rightarrow 0$. Note that exact shape of 
the isotropization function is unknown, further studies are required to determine it.

\begin{figure}
\centering
\includegraphics[width=7.6cm]{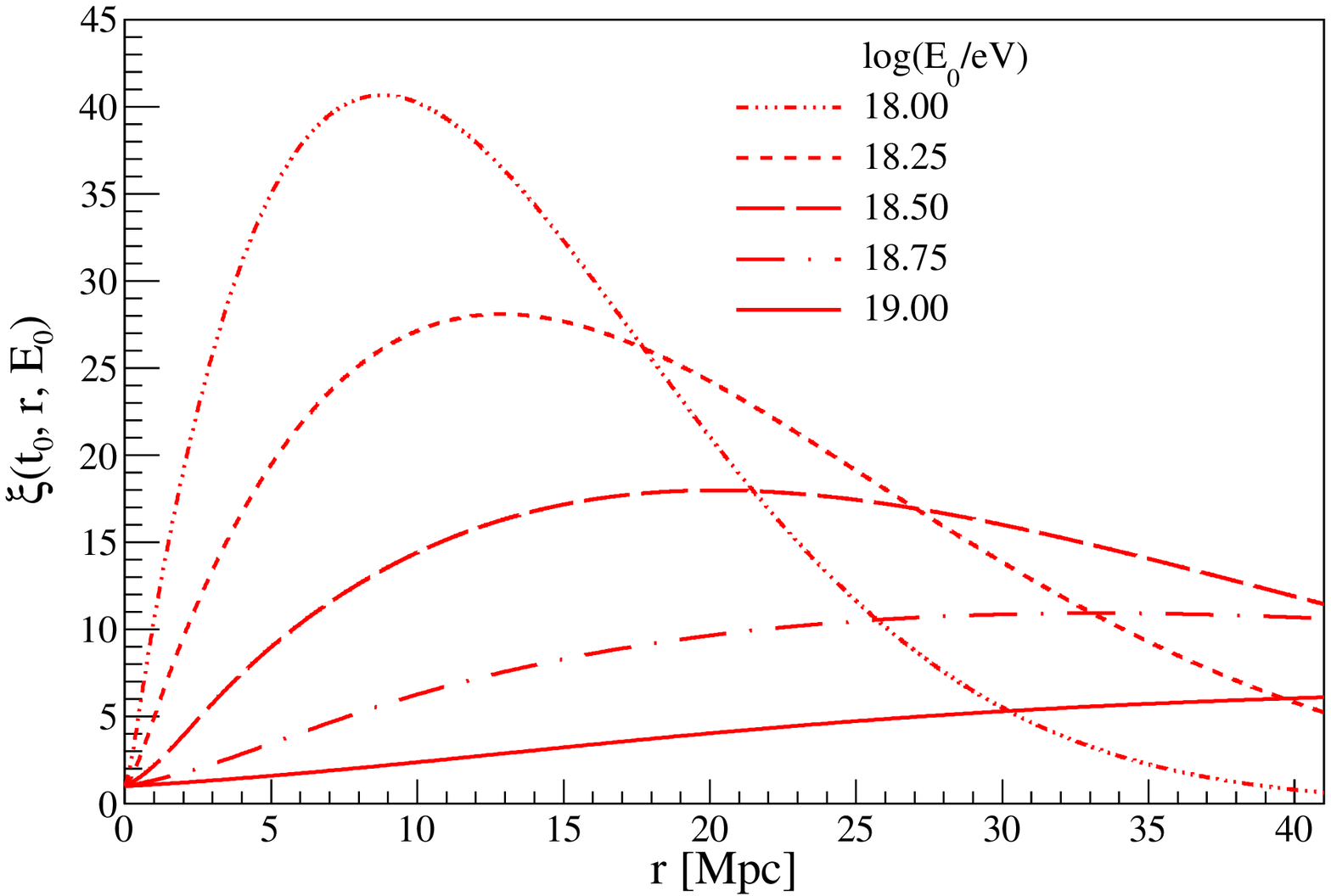}
\includegraphics[width=7.6cm]{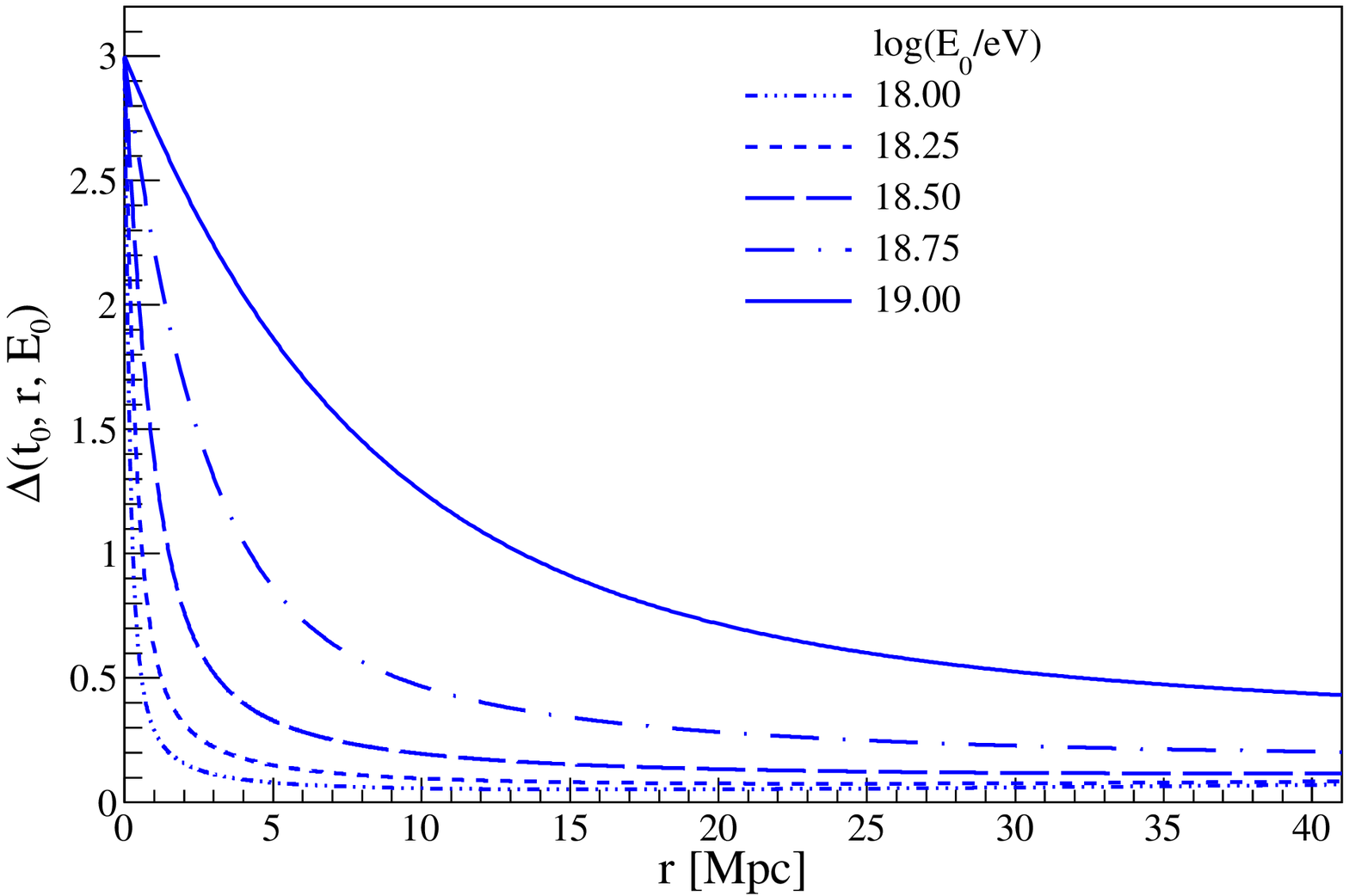}\\
\includegraphics[width=7.6cm]{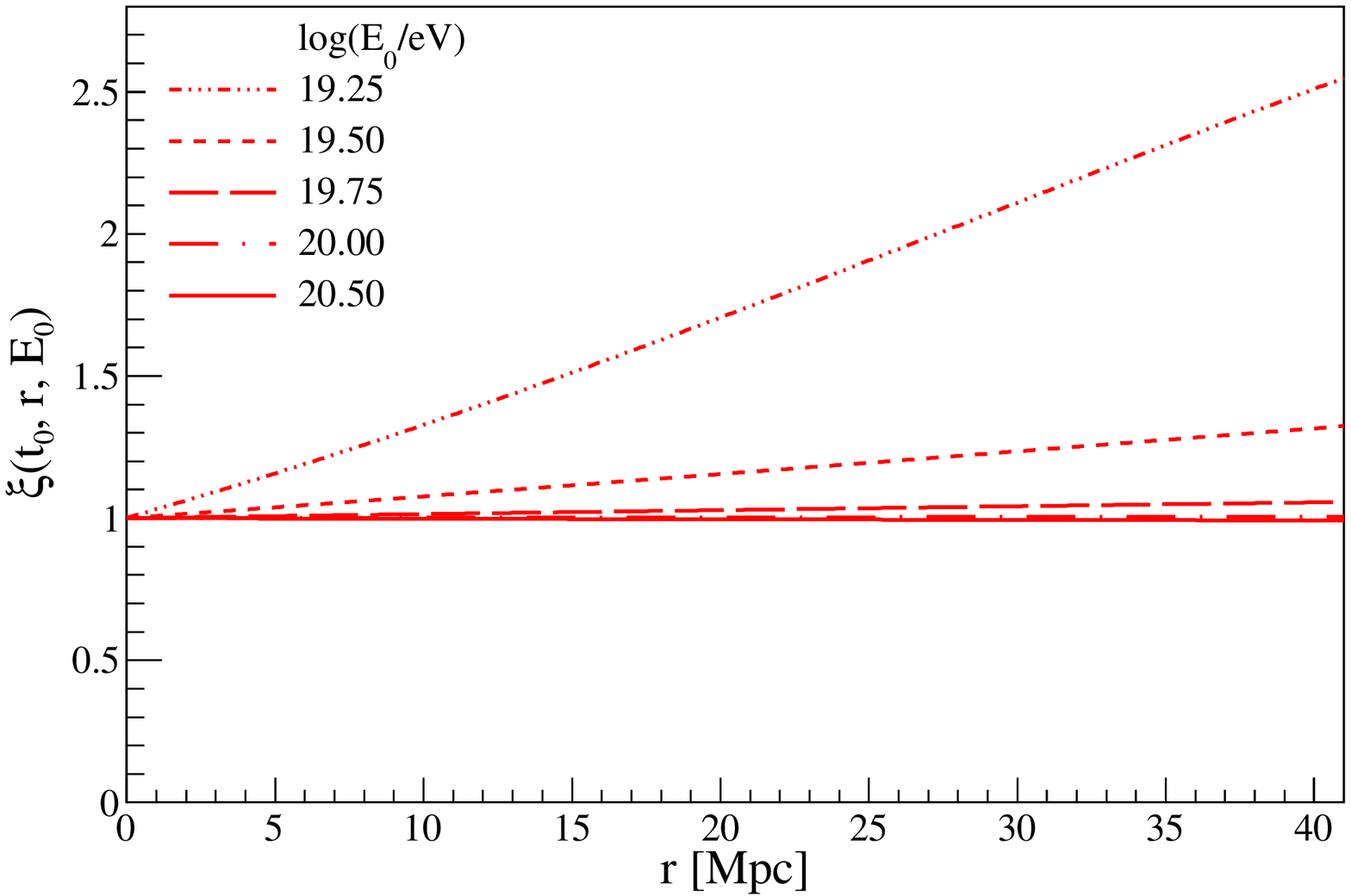}
\includegraphics[width=7.6cm]{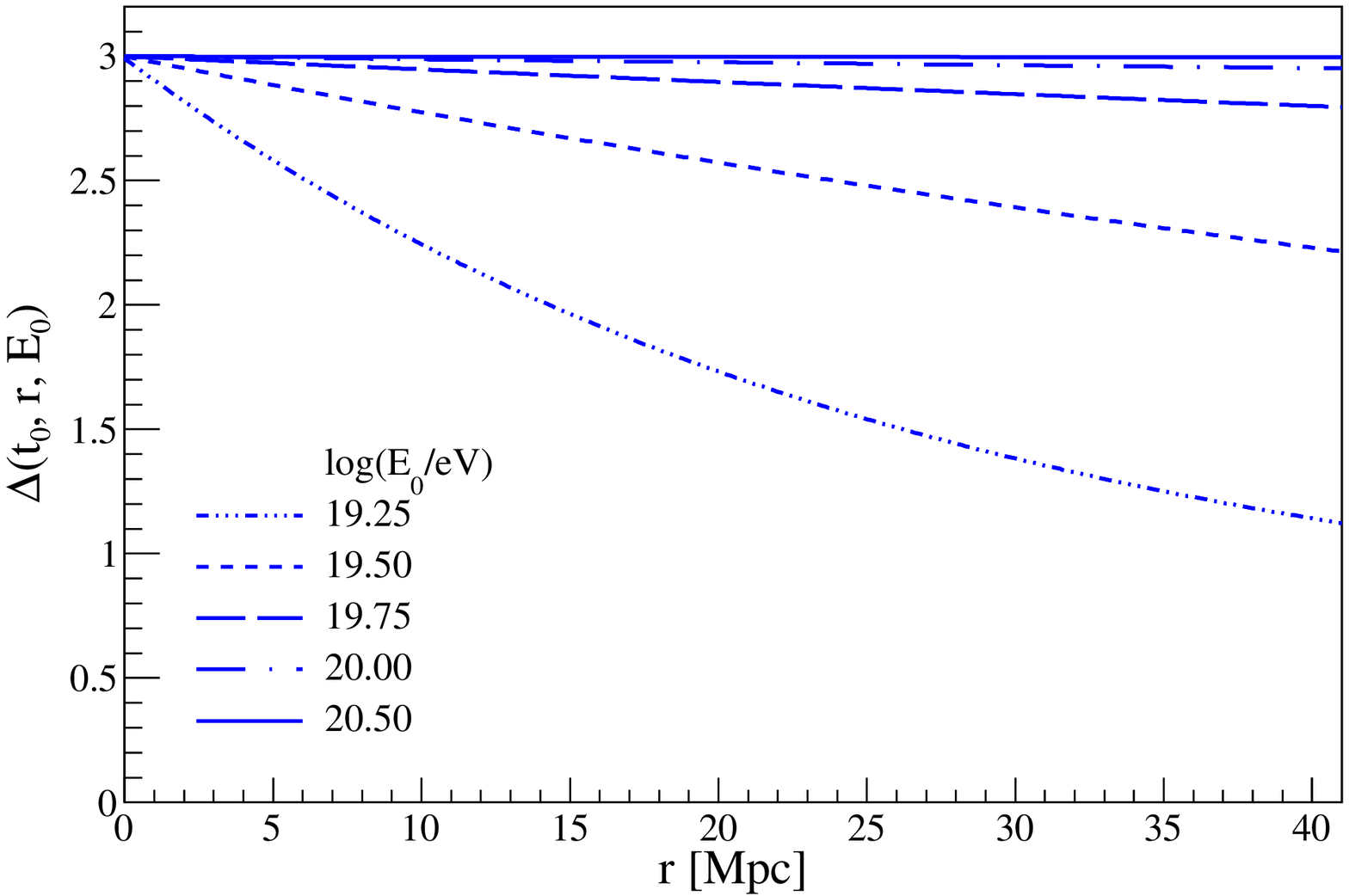}
\caption{$\xi(t_0,r,E_0)$ and $\Delta(t_0,r,E_0)$ for protons as a function of the comoving distance
$r$ for $\log(E_0/\textrm{eV})= 18,\, 18.25,\, 18.5,\, 18.75,$ and $19$ (top panels) and for
$\log(E_0/\textrm{eV}) = 19.25,\, 19.5,\, 19.75,\, 20$, and $20.5$ (bottom panels).
The paremeters consider in the calculations are: $z=0.2$, $\gamma=2$, $B=10$ nG, and
$l_c=0.5$ Mpc.  \label{XiEta}}
\end{figure}

Figure \ref{XiEta} shows $\xi(t_0,r,E_0)$ and $\Delta(t_0,r,E_0)$ as a function of $r$ for $\gamma=2$,
$t_g$ corresponding to redshift $z=0.2$, $B=10$ nG, $l_c=0.5$ Mpc, and for different values of $E_0$.
At $E_0=10^{18}$ eV the propagation is done mainly in the diffusive regime since $\tau$ ranges from
0.32 Mpc at $t=t_g$ to 0.26 Mpc at $t=t_0$. The diffusive character of the propagation can be seen at 
the upper panels of the figure. As the energy increases $\tau$ also increases in such a way that at 
$E_0=10^{20.5}$ eV, it ranges from $13500$ Mpc at $t=t_g$ to $9380$ Mpc at $t=t_0$. Therefore,
for increasing values of the energy $\tau \rightarrow \infty$ and $\phi \rightarrow 1$. These limit
values correspond to the ballistic propagation regime. As can be seen from the figure, for increasing
values of energy $\xi(t_0,r,E_0)$ and $\Delta(t_0,r,E_0)$ tend to the solution found for the ballistic 
regime of propagation which is studied in detail in Appendix \ref{AS}. This is clearly seen from the 
bottom-left panel of the figure where $\xi$ tends to one for distances relatively close to $r=0$. From 
the bottom-right panel of the figure, it can also be seen that the dipole amplitude as a function of 
$r$ tends to the constant function $\Delta(t_0,r,E_0) = 3$, which corresponds to what is expected for 
the ballistic propagation regime.

\section{Conclusions}

The effects of the turbulent intergalactic magnetic field on the propagation of the ultrahigh energy
cosmic rays can play a very important role to explain current experimental data. Motivated by this
fact, we have studied the propagation of the ultrahigh energy cosmic rays in presence of the turbulent
intergalactic magnetic field, considering the Boltzmann equation in a curved space-time, to take into
account the expansion of the Universe. Considering the moments of the distribution we have obtained a
system of partial differential equations for the number density of particles and for the flux that
take into account the ballistic and the diffusive regimes of propagation as well as the transition
between them. We have found that in both, the diffusive and ballistic limits, the known solutions
are recovered. Finally, we have solved numerically the equations system for a simplified case in
which it can be clearly seen the transition from the diffusive to the ballistic regimes of propagation
as the energy measured at a given comoving distance from the source increases.

It is worth mentioning that the equations system found can be extended in order to include the
propagation of different nuclear species with their respective interactions (photodisintegration
and photopion production) in the radiation field present in the Universe. This can be used for
the development of models of the ultrahigh energy cosmic ray flux that include the effect of the
intergalactic magnetic field.

\appendix
\section{Analytic solution for the ballistic regime of propagation}
\label{AS}

In the ballistic regime $\phi(t,r,E) \cong 1$ and $\tau(t,E)\rightarrow \infty$, then eqs.~(\ref{xir})
and~(\ref{etar}) become,
\begin{align}
\label{xiballr}
\frac{\partial \xi}{\partial t} + \frac{1}{a} \frac{\partial \eta}{\partial r} &= %
a^{\gamma-1}(t)\, \delta(r)\, \Theta(t-t_g), \\
\label{etaballr}
\frac{\partial \eta}{\partial t} + \frac{1}{a} \frac{\partial \xi}{\partial r} &= 0.
\end{align}
Eqs.~(\ref{xiballr}) and~(\ref{etaballr}) can be solved by using the method of characteristics~\cite{CharactM}. 
The solution is such that $\xi(t_0,r,E_0) = \eta(t_0,r,E_0)$, where
\begin{equation}
\label{xiint}
\xi(t_0,r,E_0) = \int_{t_g}^{t_0} dt' \, a^{\gamma-1}(t')\ \delta\left( r-%
\int_{t'}^{t_0} \frac{dt''}{a(t'')} \right)= a^\gamma\left(\widetilde{r}^{\: \text{-1}}(\widetilde{r}_0-r) \right)
\Theta(\widetilde{r}_0-r).
\end{equation}
Here,
\begin{equation}
\widetilde{r}(t) = \int_{t_g}^t \frac{dt''}{a(t'')},
\end{equation}
is the comoving distance traveled by a massless particle that is injected at $r=0$ at time $t_g$,
$\widetilde{r}^{\: \text{-1}}(x)$ is the inverse function of $\widetilde{r}(t)$, and
$\widetilde{r}_0= \widetilde{r}(t_0)$. Note that for $r\ll \widetilde{r}_0$ it follows that
$\xi(t_0,r,E_0) \cong \Theta(\widetilde{r}_0-r)$.

\begin{figure}
\centering
\includegraphics[width=7.6cm]{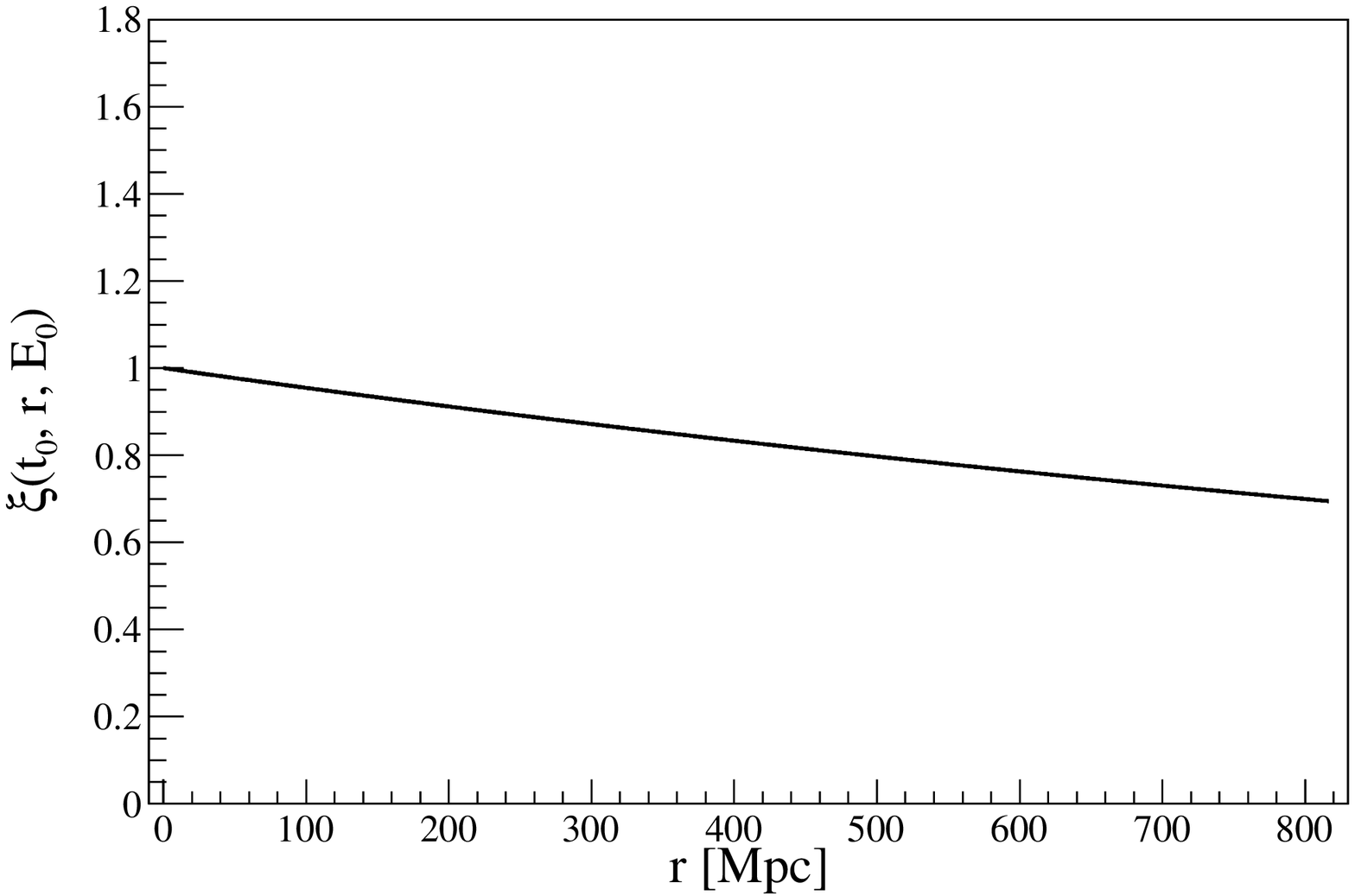}
\includegraphics[width=7.6cm]{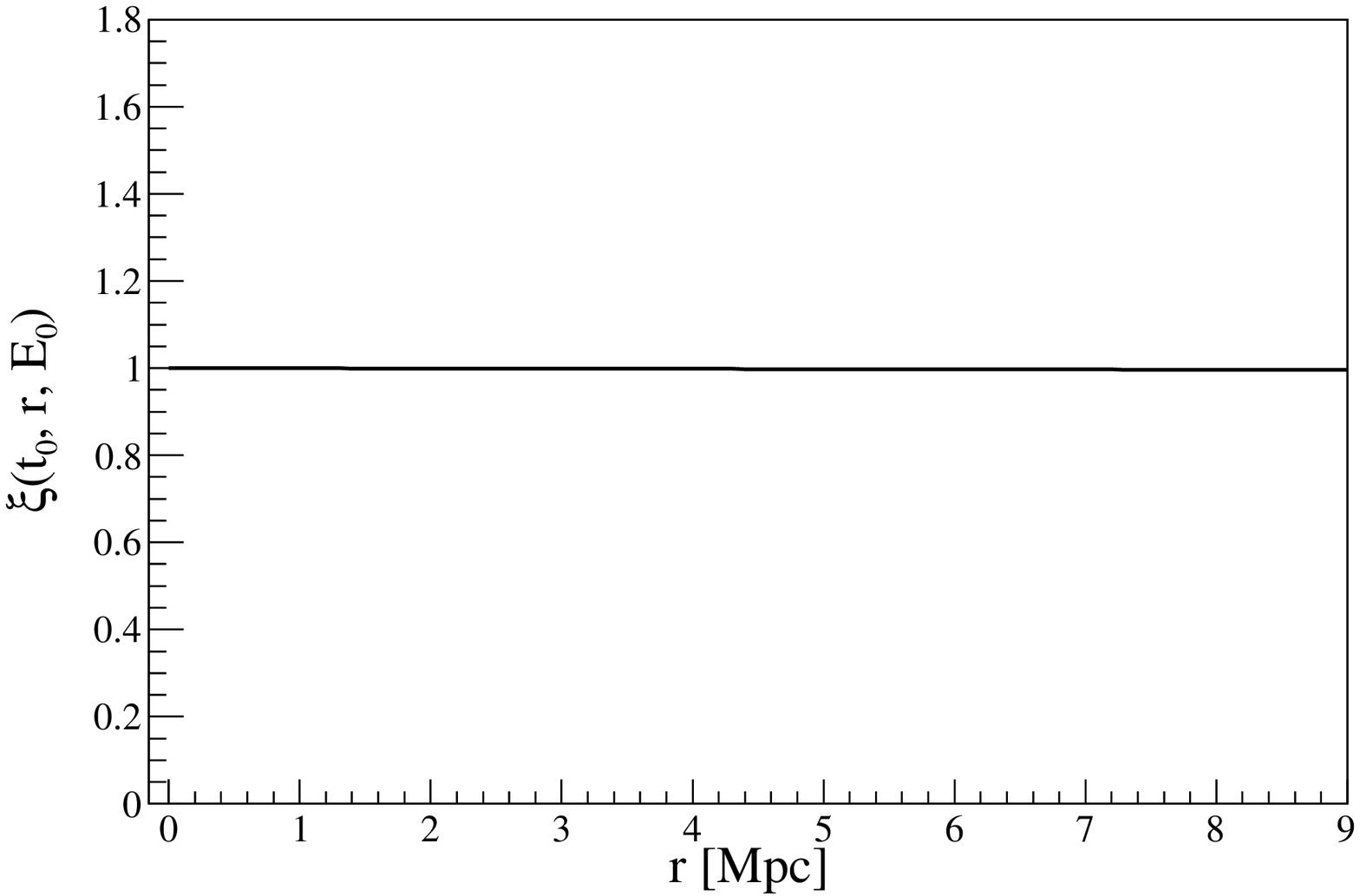}
\caption{$\xi(t_0,r,E_0)$ as a function of $r$ for protons. The right panel shows the region 
$r \ll \widetilde{r}_0$ in more detail. The spectral index and the redshift used in the
calculation are $\gamma=2$ and $z=0.2$, respectively.
\label{XiEtaBall}}
\end{figure}

The left panel of fig.~\ref{XiEtaBall} shows $\xi(t_0,r,E_0)$ as a function of $r$ from $0$ to 
$\widetilde{r}_0$ for the same parameters used in the full numerical calculation of section \ref{SE}, 
i.e.~$z=0.2$ and $\gamma=2$. From the left panel of the figure it can be seen that $\xi$ decreases 
with $r$. This is due to the fact that the protons that reach larger values of $r$ with energy $E_0$ 
at time $t_0$ have to be injected by the source with larger energies, due to the adiabatic energy 
loss undergone by them during propagation. The more energetic particles are less numerous due to 
the decrease of the energy spectrum with energy. From the right panel of the figure it can be seen 
that for $r \ll \widetilde{r}_0$ it follows that $\xi \cong 1$.

In the ballistic regime of propagation the dipole amplitude is given by eq.~(\ref{DeltaS}), which
gives $\Delta(t_0,r,E_0) = 3$ due to the fact that, in this case, $\xi(t_0,r,E_0) = \eta(t_0,r,E_0)$.

\acknowledgments

A.~D.~S.~is member of the Carrera del Investigador Cient\'ifico of CONICET, Argentina. This work is
supported by ANPCyT PICT-2015-2752, Argentina. The author thanks the members of the Pierre Auger
Collaboration, specially R. Clay for reviewing the manuscript. 


\begin{thebibliography}{99}

\bibitem{AugerTA:17}
D. Ivanov,  \emph{Report of the
  Telescope Array - Pierre Auger Observatory Working Group on Energy
  Spectrum},
\href{https://doi.org/10.22323/1.301.0498}{\emph{PoS} {\bfseries
  ICRC2017} (2018) 498}.

\bibitem{Science:17}
{\scshape Pierre Auger} collaboration,  \emph{Observation of a Large-scale
  Anisotropy in the Arrival Directions of Cosmic Rays above $8 \times 10^{18}$
  eV},
\href{https://doi.org/10.1126/science.aan4338}{\emph{Science}
  {\bfseries 357} (2017) 1266}
  [\arXivid{1709.07321}].

\bibitem{Auger:12}
{\scshape Pierre Auger} collaboration,  \emph{Large scale distribution of
  arrival directions of cosmic rays detected above $10^{18}$\,eV at the Pierre
  Auger Observatory},
\href{https://doi.org/10.1088/0067-0049/203/2/34}{\emph{Astrophys.\ J.\
  Suppl.} {\bfseries 203} (2012) 34}
  [\arXivid{1210.3736}].

\bibitem{Mollerach:19}
S.~Mollerach and E.~Roulet,  \emph{Ultrahigh energy cosmic rays from a nearby
  extragalactic source in the diffusive regime},
\href{https://doi.org/10.1103/PhysRevD.99.103010}{\emph{Phys.\ Rev.\ D}
  {\bfseries 99} (2019) 103010}
  [\arXivid{1903.05722}].

\bibitem{Mollerach:20}
S.~Mollerach and E.~Roulet,  \emph{Extragalactic cosmic rays diffusing from two
  populations of sources},
\href{https://doi.org/10.1103/PhysRevD.101.103024}{\emph{Phys.\ Rev.\ D}
  {\bfseries 101} (2020) 103024}
  [\arXivid{2004.04253}].

\bibitem{Harari:14}
D.~Harari, S.~Mollerach and E.~Roulet,  \emph{Anisotropies of ultrahigh energy
  cosmic rays diffusing from extragalactic sources},
\href{https://doi.org/10.1103/PhysRevD.89.123001}{\emph{Phys.\ Rev.\ D}
  {\bfseries 89} (2014) 123001}
  [\arXivid{1312.1366}].

\bibitem{CRPropa3:16}
R.~Alves~Batista, A.~Dundovic, M.~Erdmann, K.-H.~Kampert, D.~Kuempel,
  G.~M\"uller et~al.,  \emph{CRPropa 3 - a Public Astrophysical Simulation
  Framework for Propagating Extraterrestrial Ultra-High Energy Particles},
\jcap{05}{2016}{038} [\arXivid{1603.07142}].

\bibitem{Berezinsky:06}
V.~Berezinsky and A.Z.~Gazizov,  \emph{Diffusion of cosmic rays in expanding
  universe},
\href{https://doi.org/10.1086/502626}{\emph{Astrophys.\ J.}
  {\bfseries 643} (2006) 8}
  [\astroph{0512090}].

\bibitem{Aloisio:09}
R.~Aloisio, V.~Berezinsky and A.~Gazizov,  \emph{Superluminal problem in
  diffusion of relativistic particles and its phenomenological solution},
\href{https://doi.org/10.1088/0004-637X/693/2/1275}{\emph{Astrophys.\ J.}
  {\bfseries 693} (2009) 1275}
  [\arXivid{0805.1867}].

\bibitem{Prosekin:15}
A.Y.~Prosekin, S.R.~Kelner and F.A.~Aharonian,  \emph{On transition of
  propagation of relativistic particles from the ballistic to the diffusion
  regime},
\href{https://doi.org/10.1103/PhysRevD.92.083003}{\emph{Phys.\
  Rev.\ D} {\bfseries 92} (2015) 083003}
  [\arXivid{1506.06594}].

\bibitem{Pettinari:16}
G. Pettinari, \emph{The Intrinsic Bispectrum of the Cosmic Microwave Background}, Springer International Publishing, Cham, Switzerland (2016).

\bibitem{Berenstein:88}
J. Bernstein, \emph{Kinetic Theory in the Expanding Universe}, Cambridge University Press, Cambridge, U.K. (1988).

\bibitem{Ahlers:17}
M.~Ahlers and P.~Mertsch,  \emph{Origin of Small-Scale Anisotropies in Galactic
  Cosmic Rays},\\
\href{https://doi.org/10.1016/j.ppnp.2017.01.004}{\emph{Prog.\
  Part.\ Nucl.\ Phys.} {\bfseries 94} (2017) 184}
  [\arXivid{1612.01873}].

\bibitem{Ahlers:14}
M.~Ahlers,  \emph{Anomalous Anisotropies of Cosmic Rays from Turbulent Magnetic
  Fields},\\
\href{https://doi.org/10.1103/PhysRevLett.112.021101}{\emph{Phys.\
  Rev.\ Lett.} {\bfseries 112} (2014) 021101}
  [\arXivid{1310.5712}].

\bibitem{Deligny:19}
O.~Deligny,  \emph{Measurements and implications of cosmic ray anisotropies
  from TeV to trans-EeV energies},
\href{https://doi.org/10.1016/j.astropartphys.2018.08.005}{\emph{Astropart.\
  Phys.} {\bfseries 104} (2019) 13}
  [\arXivid{1808.03940}].

\bibitem{CharactM}
I. Stavroulakis and S. Tersian, \emph{Partial Differential Equations An Introduction with Mathematica and MAPLE (Second Edition)}, World Scientific, New Jersey, U.S.A. (2004).

\bibitem{Han:17}
J.L. Han, \emph{Observing Interstellar and Intergalactic Magnetic Fields}, \href{https://doi.org/10.1146/annurev-astro-091916-055221}{\emph{Annu.\ Rev.\ Astron.\ Astrophys.} {\bfseries 55} (2017) 111}.

\bibitem{Mollerach:13}
S.~Mollerach and E.~Roulet,  \emph{Magnetic diffusion effects on the ultra-high
  energy cosmic ray spectrum and composition},
\jcap{10}{2013}{013} [\arXivid{1305.6519}].

\bibitem{KT:94}
E. Kolb, M. Turner, \emph{The Early Universe}, Addison-Wesley, Redwood, U.S.A. (1988).

\bibitem{Thomas:95}
J. Thomas, \emph{Numerical Partial Differential Equations: Finite Difference Methods}, Springer, New York, U.S.A. (1995).

\end{thebibliography}
\end{document}